\documentclass[%
 amsmath,amssymb,
 aps,
prb,
twocolumn
]{revtex4}

\usepackage{amssymb,amsmath}
\usepackage{color}
\usepackage{natbib}
\usepackage{hyperref}
\usepackage{setspace}
\usepackage{floatrow}
\usepackage[english]{babel}
\usepackage{sidecap} 
\usepackage[T2A,OT1]{fontenc}
\usepackage[utf8]{inputenc}
\usepackage{graphicx}
\usepackage{dcolumn}
\usepackage{bm}

\begin{document}

\newcommand{\tens}[1]{\hat{#1}}
\newcommand{\refPR}[1]{[\onlinecite{#1}]}
\newcommand{\eps}{\varepsilon}
\newcommand{\kap}{\varkappa}


\title{Topological transition in coated wire medium}

\author{Maxim A. Gorlach, Mingzhao Song, Alexey P. Slobozhanyuk, Andrey A. Bogdanov, Pavel A. Belov}
\affiliation{ITMO University, Saint Petersburg 197101, Russia}


\begin{abstract}
We develop a theory of nonlocal homogenization for metamaterial consisting of parallel metallic wires with dielectric coating. It is demonstrated that manipulation of dielectric contrast between wire dielectric shell and host material results in switching of metamaterial dispersion regime from elliptic to the hyperbolic one, i.e. the topological transition takes place. We confirm our theoretical predictions by full-wave numerical simulations.  
\end{abstract}

\maketitle


\section{Introduction}

Tunable metamaterials constitute one of the promising and rapidly developing branches of metamaterial physics~[\onlinecite{Shadrivov-book,Boardman,Liu}]. Implementation of tunable structures is vital for various applications such as fabrication of metamaterial-based devices~[\onlinecite{Zheludev}] and tuning of metamaterial resonant response in a wide range~\refPR{Powell-2007,Slob2012,Shadrivov-2006,Kang,Singh,Koshelev,Kapitanova-11,Shadrivov-Kapitanova,Kapitanova-12,Lapine-2011,Slobozhanyuk,Liu-2013}. In order to achieve such tuning it was proposed to employ nonlinear elements~\refPR{Powell-2007,Slob2012}, static electric or magnetic fields~[\onlinecite{Shadrivov-2006,Kang}], temperature~[\onlinecite{Singh,Koshelev}], photosensitive elements~\refPR{Kapitanova-11,Shadrivov-Kapitanova,Kapitanova-12} or mechanical interactions~[\onlinecite{Lapine-2011,Slobozhanyuk,Liu-2013}].

One of important problems existing in the field is implementation of a tunable metamaterial that can be switched from elliptic to the hyperbolic dispersion regime. Such switching is often termed as topological transition~\refPR{Krish,Shchelokova,Gomez} because of the drastic change in the topology of isofrequency surfaces. Topological transition is also accompanied by switching from positive to negative refraction scenario along with the increase in the photonic density of states, spontaneous emission enhancement and improved superlensing effects~\refPR{Iorsh}.

\begin{figure}[b]
    \includegraphics[width=1.0\linewidth]{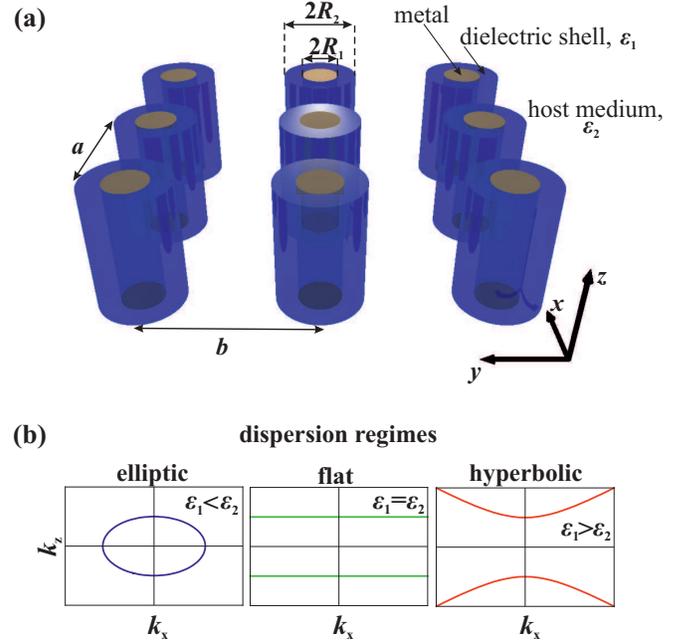}
    \caption{(a) A schematic representation of coated wire medium. (b) Typical isofrequency contours for coated wire metamaterial. Tuning the dielectric contrast between wire shell and host medium makes possible to switch between different dispersion regimes.}
    \label{ris:Core-shell}
    \end{figure}

Note that the majority of topological transitions reported previously were induced by the frequency variation. However, in our work we consider a way to realize the topological transition at {\it fixed} frequency by varying system parameters. Specifically, we investigate electromagnetic properties of metamaterial based on the array of parallel metallic wires coated with the dielectric shell with permittivity $\varepsilon_1$, arranged in the sites of rectangular lattice $a\times b$ and placed into host medium with permittivity $\varepsilon_2$ [Fig.~\ref{ris:Core-shell}(a)]. We demonstrate a simple way of switching such metamaterial from elliptic dispersion regime to the hyperbolic one by changing the  permittivity of the dielectric shell. Namely, if $\varepsilon_1<\varepsilon_2$, an elliptic dispersion regime is realized, whereas for $\varepsilon_1>\varepsilon_2$ a hyperbolic dispersion regime is observed [Fig.~\ref{ris:Core-shell}(b)]. Our findings highlight the crucial impact of the wire coating on metamaterial dispersion properties thus paving a way to the implementation of tunable `elliptic-hyperbolic' metamaterials. 

Importantly, simple wire medium composed of uncoated wires does not allow to implement such tunable metamaterial. Indeed, in the frequency region below plasma frequency simple wire medium is neither elliptic nor hyperbolic metamaterial since it possesses flat isofrequency surface corresponding to TEM mode [Fig.~\ref{ris:Core-shell}(b)]~[\onlinecite{Belov2003,Simovski-AM}]. On the other hand, in the spectral range above plasma frequency wire medium exhibits the trirefringence phenomenon due to spatial dispersion effects~\refPR{Belov2003,Gorlach2015} and therefore is not suitable for the implementation of hyperbolic metamaterial.

The rest of the paper is organized as follows. In Sec.~\ref{sec:Analysis} the analytical model to describe coated wire medium is summarized. We also compare our analytical approach with already studied limiting cases and numerical modeling. Section~\ref{sec:Tunable} demonstrates the possibility of switching coated wire medium from elliptic dispersion regime to the hyperbolic one. Our conclusion is supported both by calculated isofrequency contours and by the simulation of Gaussian beam refraction at the boundary of metamaterial. Drawn conclusions appear in Sec.~\ref{sec:Conclusion}. Calculation of the effective nonlocal permittivity tensor for coated wire medium is outlined in Appendix.

\section{Analytical model for coated wire metamaterial}\label{sec:Analysis}
It turns out that electromagnetic properties of coated wire medium can be described analytically under thin wire approximation ($R_2\ll a$). As it is shown in Appendix, effective nonlocal permittivity of coated wire medium is given by the expression
\begin{equation}\label{EffPermittivityFinal0}
\begin{split}
& \varepsilon_{zz}(q,k_z)=\varepsilon_2+\\
&\left[-\frac{\varkappa_2^2}{\varepsilon_2\,q_0^2}+\frac{ab\,\varkappa_1^2}{2\pi}\,\frac{\ln(R_1/R_2)}{\varepsilon_1+(\varepsilon_1-\varepsilon_2)\,\varkappa_1^2 R_2^2/2\,\ln(R_1/R_2)}\right]^{-1}\:,\\
\end{split}
\end{equation}
where $q=\omega/c$, $\kap_{1,2}=\sqrt{q^2\,\eps_{1,2}-k_{\rm{z}}^2}$, and $q_0$ is a plasma wavenumber of the wire array~\refPR{Belov2002}. In the case of square lattice with the period $a$
\begin{equation}\label{PlasmaFrq}
q_0^2=\frac{2\pi/a^2}{\ln\left(\frac{a}{2\pi\,R_2}\right)+0.5275}\:.
\end{equation}
Permittivity of the metamaterial in the transverse direction can be calculated as
\begin{equation}\label{TransversePermittivity0}
\varepsilon_{\bot}=\varepsilon_2+2\,\varepsilon_2\,\left[\frac{1}{f_V}\,\frac{R_2^2\,(\varepsilon_1+\varepsilon_2)+R_1^2\,(\varepsilon_1-\varepsilon_2)}{R_2^2\,(\varepsilon_1-\varepsilon_2)+R_1^2\,(\varepsilon_1+\varepsilon_2)}-1\right]^{-1}\:,
\end{equation}
where $f_V=\pi\,R_2^2/a^2$.

It is instructive to analyze some particular cases demonstrating that the derived nonlocal effective permittivity is consistent with the results of the previous studies of wire media~[\onlinecite{Belov2002,Belov2003,Silv2006,Simovski-AM}].

If $R_1=R_2$ or $\varepsilon_1\rightarrow\infty$, the structure is a simple wire medium composed of wires with the radius $R_2$ placed in the host medium with permittivity $\varepsilon_2$. Equation \eqref{EffPermittivityFinal} yields
\begin{equation}\label{PermittivityCase1}
\varepsilon_{\rm{zz}}(q,k_{\rm{z}})=\varepsilon_2-\frac{q_0^2}{q^2-k_{\rm{z}}^2/\varepsilon_2}\:,
\end{equation}
which is consistent with the results of Refs.~[\onlinecite{Belov2003,Simovski-AM}].

If $\varepsilon_1=\varepsilon_2$, the structure is again simple wire medium but with the wires of radius $R_1$. Equation \eqref{EffPermittivityFinal} yields
\begin{equation}\label{PermWire1}
\begin{split}
& \varepsilon_{zz}(q,k_{\rm{z}})=\varepsilon_2-\frac{q_1^2}{q^2-k_{\rm{z}}^2/\varepsilon_2}\:,\\
& q_1^2=\frac{2\pi/a^2}{\ln\left(\frac{a}{2\pi\,R_1}\right)+0.5275}\:,\\
\end{split}
\end{equation}
which is again consistent with Refs.~[\onlinecite{Belov2003,Simovski-AM}]. Note that this expression for plasma frequency refers to the wire medium with square lattice.

Finally, the limiting case $R_1\rightarrow 0$ corresponds to the structure composed of dielectric rods. In this case according to Eq.~\eqref{EffPermittivityFinal}
\begin{equation}\label{PermRod}
\varepsilon_{\rm{zz}}(q,k_{\rm{z}})=\varepsilon_2+\left[\frac{1}{f_V (\varepsilon_1-\varepsilon_2)}-\frac{q^2-k_{\rm{z}}^2/\varepsilon_2}{q_0^2}\right]^{-1}\:,
\end{equation}
where $f_V=\pi\,R_2^2/a^2$. Equation~\eqref{PermRod} coincides with the result of Ref.~[\onlinecite{Silv2006}].

\begin{figure}[hb]
    \includegraphics[width=1.0\linewidth]{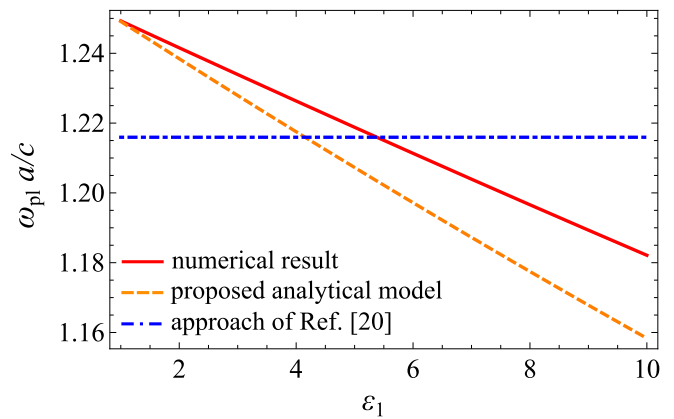}
    \caption{Plasma frequency of coated wire medium as a function of shell dielectric permittivity calculated by different approaches: dashed line for the discussed analytical model, dot-dashed curve for the approach developed in Ref.~\refPR{Tyukhtin} and solid line for the results of numerical simulation in Comsol, 2D eigenmode solution, losses in the structure are neglected. $R_1/R_2=0.05$, $R_2=0.1\,a$, $\varepsilon_2=1$.}
    \label{ris:PlasmaQ}
    \end{figure}

Furthermore, we calculate coated wire medium plasma frequency and compare our approach with the model developed in Ref.~\refPR{Tyukhtin} (Fig.~\ref{ris:PlasmaQ}). Plasma frequency is determined as frequency cutoff for TM eigenmode with the lowest frequency. In the analytical model plasma frequency is determined from the equation
\begin{equation}\label{Plasma}
\varepsilon_{\rm{zz}}(q_{\rm{pl}},0)=0\:.
\end{equation}
Whereas the model proposed in Ref.~\refPR{Tyukhtin} suggests that plasma frequency is essentially independent of shell dielectric permittivity $\varepsilon_1$, the results of simulation prove that such dependence does exist (Fig.~\ref{ris:PlasmaQ}). According to the results of numerical simulation, when shell permittivity $\eps_1$ increases from $1$ to $10$ at fixed $\eps_2=1$, $R_2=0.1\,a$ and $R_1/R_2=0.05$, plasma frequency experiences a decrease by $6\%$. Note that the developed analytical approach is in an excellent agreement with numerical simulation for low shell permittivities $\eps_1<3$. However, the agreement becomes worse for sufficiently high $\eps_1$. This can be easily understood since a wire with high permittivity $\eps_1$ of coating mimics an uncovered metallic wire with the radius $R_2$. At the same time, the validity of our approach is constrained by the thin wire approximation.

\section{Topological transition}\label{sec:Tunable}

    \begin{figure}[b]
    \includegraphics[width=1.0\linewidth]{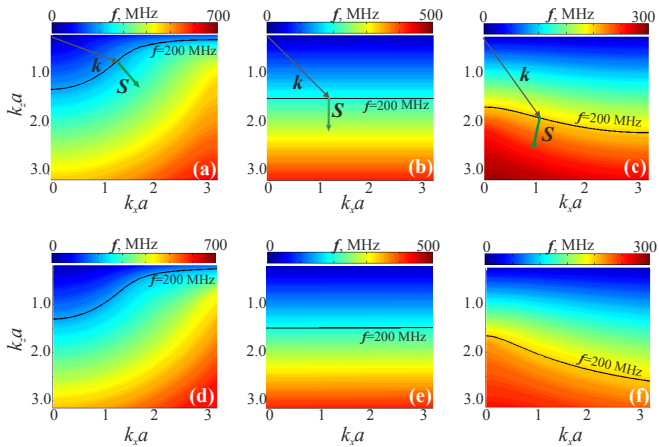}
    \caption{Calculated isofrequency contours for coated wire medium with $R_1=2.5$~mm, $R_2=10.0$~mm, $a=b=50$~mm, $\varepsilon_2=50$, $f=200$~MHz. (a-c) Numerical simulation  performed with Comsol Multiphysics software in frequency domain module using eigenmode solver. The problem was reduced to 2D one by exclusion a dependence on z-coordinate given through a factor $\exp(ik_z z)$. The simulation domain contains a single wire with Floquet boundary conditions at the unit cell boundaries. (d-f) Result of the developed analytical model. (a,d) $\varepsilon_1=1$, elliptic dispersion regime; (b,e) $\varepsilon_1=50$; (c,f) $\varepsilon_1=450$, hyperbolic dispersion regime.}
    \label{ris:IsofrPlot}
    \end{figure}

\begin{figure}[t]
    \includegraphics[width=1.0\linewidth]{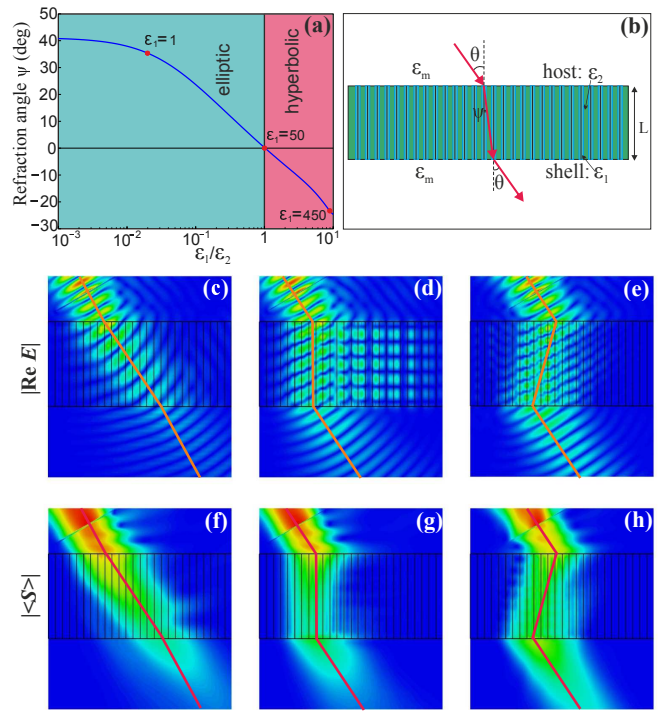}
    \caption{Different refraction scenarios for the beam incident at the boundary of coated wire medium from homogeneous medium with $\varepsilon_{\rm{m}}=81$. Wires are perpendicular to the boundary, $R_1=2.5$~mm, $R_2=10.0$~mm, $a=b=50$~mm, $\varepsilon_2=50$, $f=200$~MHz, $L=600$~mm, incidence angle is $\theta=30^{\circ}$. (a) The dependence of the refraction angle $\psi$ on dielectric contrast $\eps_1/\eps_2$ and illustration of switching between elliptic and hyperbolic dispersion regimes. (b) Schematic representation of the sample simulated numerically. (c-e) Real part of electric field; (f-h) absolute values of time-averaged energy flow. Shell dielectric permittivity is (c,f) $\varepsilon_1=1$, positive refraction; (d,g) $\varepsilon_1=50$; (e,h) $\varepsilon_1=450$, negative refraction. Orange and red lines in subfigures (c-h) are shown for eye guidance.}
    \label{ris:ES-plot}
    \end{figure}

Using the calculated effective permittivity Eqs.~\eqref{EffPermittivityFinal0} and  \eqref{TransversePermittivity0}, we analyze the properties of coated wire medium at frequencies lower then plasma frequency. In such case, there is a single propagating mode in the metamaterial, whereas at higher frequencies multiple modes can be excited. We calculate isofrequency contours for three realizations of metamaterial differing only by the permittivity of wire dielectric coating. Our analytical model as well as numerical simulation reveal that depending on the dielectric contrast between wire shell and host medium the isofrequency contour can have different curvature (Fig.~\ref{ris:IsofrPlot}), i.e. different types of the dispersion regime are realized. Figure~\ref{ris:IsofrPlot} suggests also that in the case of elliptic dispersion regime a beam incident at the boundary of metamaterial with the normal along $z$ axis will experience positive refraction whereas in the hyperbolic regime negative refraction will be observed. We confirm the latter conclusion calculating the dependence of the refraction angle on the dielectric contrast between wire shell and host medium [Fig.~\ref{ris:ES-plot}(a)] and simulating the refraction of the Gaussian beam at the boundary of coated wire medium in CST Microwave Studio. In numerical simulation, the material of the wires is annealed copper, transient solver is used and PML (perfectly matched layer)  boundary conditions are applied at all boundaries of the simulation area.

 Calculated distributions of electric field and absolute values of Poynting vector are plotted in Fig.~\ref{ris:ES-plot}(c-e) and Fig.~\ref{ris:ES-plot}(f-h), respectively. These results confirm the possibility of switching coated wire medium from elliptic dispersion regime to the hyperbolic one.

In practice shell permittivity can be varied, for instance, by changing the temperature~\refPR{Andryieuski,Zhao} or imposing external static fields~\refPR{Vendik}. In particular, it is possible to prepare host material with the permittivity weakly depending on temperature (e.g. ceramic), whereas the permittivity of the wire shell filled with water can be changed gradually from 81 to 58~\refPR{Andryieuski}. Another option is to use ceramics of two types one of which exhibits strong temperature dependence of permittivity~\refPR{Vendik}.

\section{Conclusions}\label{sec:Conclusion}
In this article, we have demonstrated a crucial impact of the wire surface on the dispersion properties of metamaterial. The properties of a wire surface can be modified by a variety of ways including bombarding by ion beam or deposition of oxide layers. In the simplest case the role of surface can be described in terms of effective coating layer.

We have developed a rigorous theoretical approach to characterize the properties of metamaterial based on coated wires. We have demonstrated that it is possible to switch coated wire medium from elliptic to the hyperbolic dispersion regime manipulating the dielectric contrast between wire shell and host medium. Our findings thus  provide a perspective to the implementation of new types of tunable metamaterials useful for various applications including engineering of new metamaterial-based devices and investigation of new physical phenomena accompanying topological transition in metamaterials.

\section{Acknowledgments}
The present work was supported by the Government of the Russian Federation (Grant No.~074-U01), a grant of the President of the Russian Federation, No.~MD-7841.2015.2, MK-6462.2016.2, Russian Foundation for Basic Research (Grant No. 15-02-08957 A, 16-37-60064, 15-32-20665) and the ``Dynasty" foundation. The authors are grateful to professors C.R.~Simovski and S.A.~Tretyakov for useful discussions. We also acknowledge Daniil Gorbach for his participation at the initial stage of the project.

\section*{Appendix. Nonlocal homogenization of coated wire medium}
In this Appendix we outline the procedure of homogenization of coated wire medium, i.e. calculation of its effective material parameters. According to the nonlocal homogenization approach~[\onlinecite{Silv2007}] the metamaterial electromagnetic properties are described in terms of the nonlocal permittivity tensor $\tens{\varepsilon}(\omega,{\bf k})$. In the subsequent derivation the CGS system of units is used and time dependence $e^{i\omega t}$ of the fields is suppressed throughout.

We consider the excitation of the structure by the external distributed sources ${\bf j}_{\rm{e}}({\bf r})={\bf j}_{\rm{e}0}\,e^{-i{\bf k}\cdot{\bf r}}$ that create the field ${\bf E}_{\rm{e}}({\bf r})={\bf E}_{\rm{e}0}\,e^{-i{\bf k}\cdot{\bf r}}$ in vacuum.  As it is known, a harmonic line current with the amplitude $I_0$ placed in host medium with permittivity $\varepsilon_2$ and oriented along $z$-axis creates the electric field with $z$-component~[\onlinecite{Felsen}]
\begin{equation}\label{Ez-one}
E_{\rm{z}}({\bf r})=-\frac{\pi\,\varkappa_2^2}{q\,c\,\varepsilon_2}\,H_0^{(2)}(\varkappa_2\,r)\,I_0\:,
\end{equation}
where $\varkappa_{1,2}=\sqrt{q^2\,\varepsilon_{1,2}-k_z^2}$, $q=\omega/c$ and $H_0^{(2)}(\kap_2\,r)$ is the Hankel function of the second kind. The current distribution in the wires of array is determined by the external excitation and reads:
\begin{equation}\label{Imn}
I_{\rm{mn}}=I_0\,e^{-ik_x\,ma-ik_y\,nb}\:.
\end{equation}
Therefore, the equation for the current in the reference wire is as follows:
\begin{equation}\label{LatticeSum}
Z\,I_0=C(q,{\bf k})\,I_0+E_{\rm{e}0}\:,
\end{equation}
where $Z$ is the wire impedance, term $E_{\rm{e}0}$ describes the external excitation and the term $C(q,{\bf k})\,I_0$ represents the local field created by all wires of the structure and acting on the reference wire. In Eq.~\eqref{LatticeSum} the lattice sum $C(q,{\bf k})$ is defined as:
\begin{equation}
C(q,{\bf k})=-\frac{\pi\,\varkappa_2^2}{q\,\varepsilon_2\,c}\,\sum\limits_{(m,n)\not=(0,0)}\,H_0^{(2)}(\varkappa_2\,r)\,e^{-ik_x\,ma-ik_y\,nb}\:.
\end{equation}
Note that the lattice sum does not depend on the structure of the wire coating. Additionally, if $C_{\rm{vac}}(q,{\bf k})$ is the lattice sum for $\varepsilon_2=1$ (vacuum), then
\begin{equation}\label{Cvac}
C(q,{\bf k})=\frac{1}{\sqrt{\varepsilon_2}}\,C_{\rm{vac}}(q,{\bf k})(q\,\sqrt{\varepsilon_2},{\bf k})\:.
\end{equation}
Effective algorithm of the sum $C_{\rm{vac}}(q,{\bf k})$ calculation was developed in Ref.~[\onlinecite{Belov2002}]. In order to determine effective  permittivity of the structure, one has to relate the electric field averaged over the unit cell to the averaged polarization. The averaging procedure is defined as follows~[\onlinecite{Alu-2011}]:
\begin{equation}\label{Average}
\left<{\bf E}\right>=\frac{1}{a\,b}\,\int\limits_{S_0}\,{\bf E}({\bf r})\,e^{i{\bf k}\cdot{\bf r}}\,dS\:,
\end{equation}
where $S_0=a\,b$ is the unit cell area. The similar expression is valid for the average polarization. Note that with this definition $\left<{\bf E}_{\rm{e}}\right>={\bf E}_{\rm{e}0}$. Equation \eqref{LatticeSum} yields:
\begin{equation}\label{Ee0}
\begin{split}
& \left<E_{\rm{ez}}\right>=\left[Z-C(q,{\bf k})\right]\,I_0=\\
& \left[Z-C(q,{\bf k})\right]\,i\,c\,q\,S_0\,\left<P_{\rm{z}}\right>\:,
\end{split}
\end{equation}
where $\left<\bf{P}\right>$ is polarization arising in addition to that induced in host medium. The field ${\bf E}_{\rm{s}}$ created by the polarized structure is related to the structure average polarization via~\refPR{Alu-2011}
\begin{equation}\label{Es}
\left<E_{\rm{sz}}\right>=-\frac{4\,\pi}{\varepsilon_2}\,\frac{\varkappa_2^2}{\varepsilon_2\,q^2-k^2}\,\left<P_{\rm{z}}\right>\:.
\end{equation}
The total field is 
\begin{equation}
\begin{split}
& \left<{\bf E}\right>=\left<{\bf E_{\rm{e}}}\right>+\left<{\bf E_{\rm{s}}}\right>=\\
& \left[Z-S(q,{\bf k})\right]\,i\,c\,q\,a\,b\,\left<\bf{P}\right>\:.
\end{split}
\end{equation}
Here,
\begin{equation}
S(q,{\bf k})=C(q,{\bf k})-\frac{4\pi i}{c\,q\,\varepsilon_2\,a\,b}\,\frac{\varkappa_2^2}{\varepsilon_2\,q^2-k^2}
\end{equation} 
is the interaction constant. On the other hand, by definition $\left<P_{\rm{z}}\right>=(\varepsilon_{\rm{zz}}(q,{\bf k})-\varepsilon_2)/(4\pi)\,\left<E_{\rm{z}}\right>$. Thus, we derive the $\rm{zz}$-component of the effective permittivity tensor:
\begin{equation}\label{EffPermittivity}
\varepsilon_{\rm{zz}}(q,{\bf k})=\varepsilon_2+\frac{4\pi i}{q\,c\,a\,b}\,\left[S(q,{\bf k})-Z\right]^{-1}\:.
\end{equation}
In Eq.~\eqref{EffPermittivity} $z$ axis is assumed to be parallel to the wires. This expression is fully consistent with that derived in Ref.~[\onlinecite{Silv2006}] for the structure composed of dielectric rods. The only difference between these expressions is in the value of impedance $Z$. It should be emphasized that the obtained result is valid only under thin wire approximation $r_0\ll{\rm min}(a,b)$ since in this approach the field of a single wire is described as a field of line current [see Eq.~\eqref{Ez-one}].

Now we discuss the calculation of the coated wire impedance. The impedance $Z$ is defined as the ratio of local field acting on a wire to the amplitude of current flowing in this wire: $Z=E^{\rm{loc}}_{\rm{z}}/I_0$. The current amplitude $I_0=I_{\rm{wire}}+I_{\rm{shell}}$ includes both conduction current in the metallic wire $I_{\rm{wire}}$ and displacement current $I_{\rm{shell}}$ in the dielectric shell. Under a monochromatic excitation ${\bf E}_{\rm{e}}({\bf r})={\bf E}_{\rm{e}0}\,e^{-i\,{\bf k}\cdot{\bf r}}$ the distribution of the fields inside and outside of the wire is described by the formulas:
\begin{equation}\label{Ez0}
E_{\rm{z}}(r)=0 \mspace{10mu} \text{for } r<R_1\:,
\end{equation}
\begin{equation}\label{Ez1}
\begin{split}
E_{\rm{z}}(r)=\sum\limits_{n=-\infty}^{\infty}\,\left[a_{\rm{n}}\,J_{|\rm{n}|}(\varkappa_1\,r)+b_{\rm{n}}\,N_{|\rm{n}|}(\varkappa_1\,r)\right]\,e^{in\varphi}\\
\text{for } R_1<r<R_2\:,
\end{split}
\end{equation}
\begin{equation}\label{Ez2}
\begin{split}
E_{\rm{z}}(r)=\sum\limits_{n=-\infty}^{\infty}\,c_{\rm{n}}\,H_{|\rm{n}|}^{(2)}(\varkappa_2\,r)\,e^{in\varphi}+E_{\rm{e}}(r)\\
\text{for } r>R_2\:,
\end{split}
\end{equation}
\begin{equation}
E_{\rm{e}}(r)\equiv E_{\rm{e}0}\,e^{-i\,{\bf k}\cdot{\bf r}}=\sum\limits_{n=-\infty}^{\infty}\,(-i)^n\,E_{\rm{e}0}\,J_{|\rm{n}|}(\varkappa_2\,r)\,e^{i\,n\varphi}\:.\label{Eext}
\end{equation}
Here $J_{\rm{n}}(\kap_2\,r)$, $N_{\rm{n}}(\kap_2\,r)$ and $H_{\rm{n}}^{(2)}(\kap_2\,r)$ are $n$-th order Bessel function, Neumann function and Hankel function of the second kind, respectively. The expansion coefficients are determined from the boundary conditions
\begin{gather}
E_{z}(R_1+0)=0\:,\label{Bound1}\\
E_{z}(R_2-0)=E_z(R_2+0)\:,\label{Bound2}\\
H_\varphi(R_2-0)=H_\varphi(R_2+0)\:,\label{Bound3}
\end{gather}
where $H_\varphi(r)=-i\,q\,\varepsilon_{\rm{p}}/\varkappa_{\rm{p}}^2\,\partial\,E_z/\partial\,r$ with $p=1$ in the area $R_1<r<R_2$ and $p=2$ in the area $r>R_2$. Making use of thin wire approximation we truncate the series in Eqs.~\eqref{Ez0}-\eqref{Eext} leaving only terms with $n=0$. Then only three unknown  coefficients $a_0$, $b_0$ and $c_0$ remain. Their values are determined from the boundary conditions Eqs.~\eqref{Bound1}-\eqref{Bound3}. Current in the coated wire can be expressed in terms of the fields as follows:

\begin{equation}\label{Iwire}
I_{\rm{wire}}=c\,R_1/2\,B_\varphi(r)\:,
\end{equation}
\begin{equation}\label{Ishell}
I_{\rm{shell}}=\int\limits_{R_1}^{R_2}\,i\,q\,c\,(\varepsilon_1-\varepsilon_2)/(4\pi)\,E_z(r)\,2\pi\,r\,dr\:.
\end{equation}

Calculating the total current by Eqs.~\eqref{Iwire}, \eqref{Ishell} and using the definition of impedance, we finally derive the following expression:

\begin{equation}\label{CoatedWireImpedance}
\begin{split}
&Z=\frac{\pi\,\varkappa_2^2}{q\,c\,\varepsilon_2}-\frac{2i\,\varkappa_2^2}{qc\,\varepsilon_2}\times\\
&\left\lbrace\ln\frac{\varkappa_2\,R_2}{2}+\gamma_e+\frac{\varkappa_1^2}{\varkappa_2^2}\,\frac{\varepsilon_2\,\ln (R_1/R_2)}{\varepsilon_1+(\varepsilon_1-\varepsilon_2)\,\varkappa_1^2\,R_1^2/2\,\ln (R_1/R_2)}\right\rbrace\:.
\end{split}
\end{equation}
Here, $\varkappa_{1,2}=\sqrt{q^2\,\varepsilon_{1,2}-k_z^2}$, $\gamma_e\approx~0.5772$ is the Euler's constant and $q=\omega/c$. Note that the real part of the wire impedance is responsible for the radiation loss contribution and it satisfies the condition ${\rm Re}\,Z={\rm Re}\,C(q,{\bf k})$ for any real-valued $q$ and ${\bf k}$. Substituting wire impedance calculated from Eq.~\eqref{CoatedWireImpedance} into Eq.~\eqref{EffPermittivity} and using the approximate formula~[\onlinecite{Belov2002,Belov2003}] for the interaction constant, we finally obtain
\begin{equation}\label{EffPermittivityFinal}
\begin{split}
& \varepsilon_{zz}(q,k_z)=\varepsilon_2+\\
&\left[-\frac{\varkappa_2^2}{\varepsilon_2\,q_0^2}+\frac{ab\,\varkappa_1^2}{2\pi}\,\frac{\ln(R_1/R_2)}{\varepsilon_1+(\varepsilon_1-\varepsilon_2)\,\varkappa_1^2 R_2^2/2\,\ln(R_1/R_2)}\right]^{-1}\:,\\
\end{split}
\end{equation}
where plasma wavenumber of the wire array $q_0$ is determined in Ref.~\refPR{Belov2003}. Since the wires are thin, their polarization along $x$ and $y$ axes is sufficiently small and the polarizability can be calculated by means of quasistatic formulas. Assuming a square lattice, we conclude that $\varepsilon_{xx}=\varepsilon_{yy}=\varepsilon_{\bot}$. It can be easily shown that the polarizability of a coated wire per unit length in the transverse direction is equal to
\begin{equation}\label{Polarizability}
\alpha=\frac{\varepsilon_2\,R_2^2}{2}\,\frac{R_2^2\,(\varepsilon_1-\varepsilon_2)+R_1^2\,(\varepsilon_1+\varepsilon_2)}{R_2^2\,(\varepsilon_1+\varepsilon_2)+R_1^2\,(\varepsilon_1-\varepsilon_2)}\:.
\end{equation}
Taking into account that the static interaction constant for this case is equal to $S_{\rm{s}}=2\pi/(a^2\,\varepsilon_2)$~[\onlinecite{Silv2006}], we obtain the expression for the structure permittivity in transverse direction
\begin{equation}\label{TransversePermittivity}
\varepsilon_{\bot}=\varepsilon_2+2\,\varepsilon_2\,\left[\frac{1}{f_V}\,\frac{R_2^2\,(\varepsilon_1+\varepsilon_2)+R_1^2\,(\varepsilon_1-\varepsilon_2)}{R_2^2\,(\varepsilon_1-\varepsilon_2)+R_1^2\,(\varepsilon_1+\varepsilon_2)}-1\right]^{-1}\:,
\end{equation}
where $f_V=\pi\,R_2^2/a^2$.

\bibliography{Bibliography-wires}
\end{document}